\documentstyle[11pt, aaspp4, psfig]{article}

\def\simless{\mathbin{\lower 3pt\hbox
     {$\rlap{\raise 5pt\hbox{$\char'074$}}\mathchar"7218$}}}   %< or of order
\def\simmore{\mathbin{\lower 3pt\hbox
     {$\rlap{\raise 5pt\hbox{$\char'076$}}\mathchar"7218$}}}   %> or of order

\begin{document}

\title{Implications of the PSR~1257+12 Planetary System for
Isolated Millisecond Pulsars}

\author{M. Coleman Miller and Douglas P. Hamilton}
\affil{Department of Astronomy, University of Maryland\\
       College Park, MD  20742-2421\\
       miller@astro.umd.edu,hamilton@astro.umd.edu}

\begin{abstract}

The first extrasolar planets were discovered in 1992 around the millisecond
pulsar PSR~1257+12.  We show that recent developments in the study of
accretion onto magnetized stars, plus the existence of the innermost,
moon-sized planet in the PSR~1257+12 system, suggest that
the pulsar was born with approximately its
current rotation frequency and magnetic moment.  If so, 
this has important implications for the formation and evolution of neutron 
star magnetic fields as well as for the formation of planets around pulsars.
In particular, it suggests that some and perhaps all isolated millisecond
pulsars may have been born with high spin rates and low magnetic fields
instead of having been recycled by accretion.

\end{abstract}


\section{Introduction}

The remarkably stable rotation frequencies of millisecond pulsars (MSP), with
typical period derivatives of $10^{-21}-10^{-19}$~s~s$^{-1}$, make them
extremely sensitive to periodic perturbations such as those produced by
orbiting companions.  This sensitivity led to the first discovery of
extrasolar planets, around the Galactic disk pulsar
PSR~1257+12 (Wolszczan \& Frail  1992).  This
pulsar has a period of $P=6.219\times 10^{-3}$~s and a period derivative
of ${\dot P}=1.2\times 10^{-19}$, which in a standard magnetic dipole spindown
model implies a dipole magnetic field of
$B=3\times 10^{19}(P{\dot P})^{1/2}\,{\rm G} \approx 8.8\times 10^8$~G and
a characteristic age $\tau_c=P/2{\dot P}=8\times 10^8$~yr.
Timing residuals from PSR~1257+12
suggested that there were two  Earth-mass planets around this pulsar
(Wolszczan \& Frail 1992).  The planetary origin of the timing
residuals was confirmed by observations  (Wolszczan 1994) of the expected
secular perturbations due to interaction between these planets (Rasio et
al. 1993; Malhotra 1993; Peale 1993).  These observations also revealed
the presence of a third, moon-sized planet closer in (Wolszczan 1994).
In order of increasing distance from the pulsar, the masses and semimajor
axes of the three planets are $M_1=0.015/\sin i_1\,M_\oplus$ at 0.19~AU,
$M_2=3.4/\sin i_2\,M_\oplus$ at 0.36~AU, and $M_3=2.8/\sin i_3\,M_\oplus$
at 0.47~AU, where $i_1$, $i_2$, and $i_3$ are the orbital inclination
angles ($i=0$ is face-on, $i=90^\circ$ is edge-on).  
All three planets are in nearly circular orbits, with
eccentricities $<$0.02.  There is some evidence for a fourth planet
(Wolszczan 1996) which, if it exists,  has a mass $M_4\sim
0.05-81\,M_\oplus$ and a semimajor axis $\sim$6-29~AU, where the mass
uncertainty is largely due to uncertainty in the fractional contribution
of such a planet to the observed spindown of the pulsar (Wolszczan et al.
2000a). Recently, some doubt was cast on the existence of the innermost
planet by Scherer et al.\ (1997), who pointed out that its 25.3~day
orbital period is close to the solar rotation period at the 17$^\circ$
solar latitude of PSR~1257+12, and suggested that the modulation might
actually be due to modulation in the electron density of the solar wind in
that direction.  However, if this effect is important it would also be
expected to be observed in other millisecond pulsars.  More importantly,
the oscillation amplitude does not depend on the
radio frequency (Wolszczan et al.~2000b), contrary to what is expected for 
a plasma effect.  Hence, the 25 day
modulation of the frequency from PSR~1257+12 is due to a planet.

The existence of this system has produced much speculation about its
origin.  As discussed by Phinney \& Hansen (1993), the proposed
formation mechanisms can be divided into presupernova scenarios, in
which the planets existed before there was a neutron star in the system,
and postsupernova scenarios, in which the planets formed
after the supernova.  Podsiadlowski (1993) reviewed a large number of
these proposed mechanisms.  Presupernova scenarios include those in which
planets survive the supernova (Bailes, Lyne, \& Shemar 1991) or are
captured into orbit around the neutron star by a direct stellar collision
(Podsiadlowski, Pringle, \& Rees 1991) or are formed in orbit around
a massive binary (Wijers et al.\ 1992).  
As reviewed by Podsiadlowski (1993), all of these
mechanisms are met with serious objections.  For example, direct
stellar collisions in the Galactic disk are expected to be exceedingly
rare, and a supernova explosion in a single-star system would be
highly likely to unbind any planets initially in orbit around it
and any remaining planets would have high eccentricities.

For these reasons, more attention has focused on postsupernova scenarios.
Some models propose that PSR~1257+12 is a ``recycled"
pulsar which has been spun up by accretion, in analogy with other
millisecond pulsars.  In these models the star might have accreted matter
from a remnant disk, for example from the disrupted
remains of a merger between two white dwarfs or a white dwarf and a
neutron star (Podsiadlowski et al.\ 1991), or from a massive disk
left over from a phase of Be binary mass transfer (Fabian \& Podsiadlowski
1991), or by deflation of a Thorne-Zytkow object (Podsiadlowski et al.\ 
1991).  Alternately, the accretion could have been from a 
stellar companion, which was then removed or disguised as a planet.
One picture, which was motivated by a report of a single
planet around PSR~1829--10 that was later retracted (Bailes et al. 1991), 
is that the companion 
was evaporated by flux from the neutron star until it had planetary mass
(Bailes et al.\ 1991; Krolik 1991; Rasio, Shapiro, \& Teukolsky 1992).
A variant of this model is that as the companion is ablated it expands,
eventually being disrupted and forming a $\sim 0.1\,M_\odot$ disk around
the neutron star, from which the planets eventually form (Stevens, Rees,
\& Podsiadlowski 1992).  Another possibility is that the ablated matter
may not escape the system, instead forming a circumbinary disk from
which planets form (Tavani \& Brookshaw 1992; Banit et al.\ 1993).  The
stellar companion would, in this scenario, be evaporated completely by the 
neutron star.

A third class of postsupernova models, distinct from those involving
disks or disrupted companions, suggests that
the planets formed from fallback matter from the supernova (Bailes et
al.\ 1991; Lin, Woosley, \& Bodenheimer 1991), or from matter that
had been ablated from a binary companion prior to the supernova
(Nakamura \& Piran 1991).  The matter from which the planets formed
might also have been accreted from the companion if supernova recoil
sent the neutron star through the companion (suggested by C. Thompson;
summarized in Phinney \& Hansen 1993).  In such models the neutron star 
was born with approximately its current spin rate and 
magnetic moment, and was therefore not spun up by accretion, in contrast
to the standard formation scenario for millisecond pulsars.

Here we argue in favor of this third class of models.  We therefore
suggest that many or all isolated MSP may have
simply been born as they are now.  This alleviates potential problems with the
birthrate of MSP versus the birthrate of low-mass X-ray
binaries (LMXB; these are usually considered the progenitors of all MSP
in the recycling scenario, and in our model are still the progenitors
of binary MSP).  This picture also suggests, but does not 
require, that supernovae may produce a bimodal distribution of neutron 
star magnetic fields and possibly spin rates.

We begin our argument in \S~2 by showing that in most scenarios the
planets must have formed in approximately their current location.  We
then show that models requiring the neutron star to be spun up by
accretion subsequent to the supernova typically run into at least one
of the following problems: (1)~if the planets form before or during
the spinup, they will be evaporated by the accretion luminosity, and
(2)~if the planets form after spinup, they must form from some remnant
disk, but the particle luminosity from the neutron star is sufficient
to disperse a tenuous disk of material faster than it can be supplied
by, e.g., evaporated material from a companion.  We also argue that
the lack of planetary bodies with masses greater than Ceres around
other isolated millisecond pulsars (Wolszczan 1999) strongly
constrains the formation of this system, and in particular suggests a
probabilistic scenario in which isolated MSP either capture enough
mass to form planets or capture virtually no mass, rather than a
smooth distribution in between.  In \S~3 we summarize the allowed
formation histories, and we discuss the implications of such a
formation scenario for the MSP population in \S~4.  We present our
conclusions in \S~5.

\section{Physical Constraints on Models}

It is useful to consider first the evolutionary path leading to the
current high rate of spin of the pulsar.  Clearly, it was either born
with and has sustained a high spin frequency, or it was spun up by
accretion at some point in its evolution.  The accretion scenarios can
be further subdivided according to whether the planets formed before,
during, or after the spin-up phase, and whether they formed at their
current locations or they formed farther out and later migrated
inwards.  In this section we therefore first consider dynamical
migration, then investigate the effects of photon and particle
luminosity on gas and planetesimals in a disk.

\subsection{Dynamical Migration}

In the section following this one, we will put strong constraints on
possible planetary formation mechanisms by assuming that the planets
formed at their current distances from the pulsar.  Here, we
examine the validity of this assumption by considering possible
mechanisms by which the planets may have formed further from the
pulsar and subsequently migrated inward to their present positions. In
order for such migration to have occurred, the planets must have
interacted with an amount of mass roughly comparable to their
own. There are three main possibilities: 1) gravitational scattering
by other planets or protoplanets, 2) interaction with a disk of
planetesimals, and 3) interaction with a gas disk.

Gravitational scattering of protoplanets can be quickly ruled out as a
substantial source of radial migration for the pulsar planets under
consideration. In this scenario, we model interactions between planets
as randomly-oriented velocity impulses, and assume that several to
several tens of scattering events have happened during the formation
of the system. Large stochastic eccentricities and inclinations are
expected in systems in which significant gravitational scattering has
occurred.  For example, if cumulative scattering events are capable of
changing planetary semimajor axes by a few tens of percent, they will
also be strong enough to induce orbital eccentricities of order 0.2
and inclinations above 10 degrees. While large inclinations cannot be
ruled out in the PSR 1257+12 system, the near circular orbits of all
three planets argue strongly against significant gravitational
scattering.

Interactions with a disk of gas or planetesimals are required to
attain the nearly circular orbits that are observed.  Can these
interactions also cause planets to migrate significantly?  In the case
of planetesimals in the PSR 1257+12 system, the answer is again no.
Escape velocities from Earth-sized planets are of the order of 15~km/s
while the orbital velocity of even the outermost pulsar planet is of
order 50 km/s.  Since orbital velocities dominate escape velocities,
gravitational scattering of planetesimals is relatively weak. Even the
outermost planet cannot eject planetesimals from the system unless an
unlikely sequence of multiple favorable close approaches occurs;
collisions between planets and planetesimals are much more likely. The
planets, therefore, absorb the vast majority of the planetesimals
while nearly conserving the total angular momentum of the system. The
most that can happen is that the inner and outer planets separate
somewhat; there is no systematic inward migration.  In this scenario,
conservation of angular momentum implies that the three planets could
not all have formed further from the pulsar than they are now, thereby
avoiding the destructive mechanisms that we discuss below.

Finally, interactions with a gas disk can cause solid objects to move
inward toward the pulsar. So that angular momentum is conserved, an
equivalent amount of gas must be offset outward. If the composition of
the gas is roughly solar, the solids that condense to form the planets
account for at most a few percent of the total mass. The rest of the
mass remains in gaseous form in an extended thickened disk encircling
the pulsar. Planet-sized objects raise waves in the disk, and the
gravitational perturbations of these waves put torques back on the
planets which can systematically change their orbits. The details of
how this occurs depend most strongly on the mass of the planet, and
the radial density profile of the disk. Given a large disk mass,
planetary migration by this mechanism could be substantial.  However,
if the pulsar is to be spun up by accretion, the $\sim 10^7 - 10^8$
year spinup timescale is long compared to the expected $\sim
10^6-10^7$~yr survival time of protoplanetary disks (Bachiller 1996).
We expect that the lifetime of a protoplanetary disk will be
especially short in the high-luminosity environment of the pulsar.
Therefore, for most of the spinup time, the planets must have been
unshielded by a disk and close to their present distances.

\subsection{Accretion Luminosity and Ablation}

If the planets formed at approximately their current locations, they
could be affected by the photon or particle luminosity from the
neutron star, either during accretion or after.  Here we consider
ablation of the planets by the photon luminosity produced by
accretion, and in the next section we discuss ablation of a
protoplanetary disk and planetesimals by high-energy particles
produced by pulsar spindown.

Higher luminosity during accretion means more rapid and effective
ablation of the planets, so in order to be conservative we will calculate
the minimum luminosity required for spinup.  The luminosity, in turn, 
is related to the accretion rate, which may be estimated by magnetic 
torque balance arguments
(e.g., Ghosh \& Lamb 1979).  These arguments show that if the star is
spun up entirely by accretion (the standard assumption in LMXB recycling
scenarios), then the equilibrium spin frequency, at which the net torque
vanishes, can be characterized by the orbital frequency at some radius
$r_t$.  The required accretion rate increases rapidly with decreasing
$r_t$ and hence with increasing stellar spin frequency, so the most
conservative assumption is that the neutron star is currently spinning
at the highest frequency it has ever had.  In reality, substantial spindown
via magnetic dipole braking has likely taken place.

To calculate the luminosity required for a given $r_t$, we
write $r_t=\omega_c r_A$,
where $\omega_c$ is the ``fastness parameter" and $r_A$, the Alfv\'en
radius, is (see, e.g., Shapiro \& Teukolsky 1983, pg. 451)
\begin{equation}
r_A=3.5\times 10^8 L_{37}^{-2/7}\mu_{30}^{4/7}
\left(M\over{M_\odot}\right)^{1/7}R_6^{-2/7}\,{\rm cm.}
\end{equation}
Several recent analyses (e.g., Li \& Wang
1999; Psaltis et al.\ 1999) have
concluded that $\omega_c>0.8$.  Therefore,
\begin{equation}
r_t=2.8\times 10^8 (\omega_c/0.8)L_{37}^{-2/7}\mu_{30}^{4/7}
\left(M\over{M_\odot}\right)^{1/7}R_6^{-2/7}\,{\rm cm.}
\end{equation}
Here $L=10^{37}L_{37}$~erg~s$^{-1}$, $\mu=10^{30}\mu_{30}$~G~cm$^3$,
and $R=10^6R_6$~cm.
From this equation we see that higher $\omega_c$ means
larger $r_t$ for a given luminosity, implying a lower stellar
spin frequency.  Therefore, the required luminosity increases with
increasing $\omega_c$.  The equilibrium period is then
$P_{\rm eq}=2\pi(r_t^3/GM)^{1/2}$ (valid in Schwarzschild spacetime).  
This period cannot exceed the
current rotational period of PSR~1257+12, $P=6.2\times 10^{-3}$~s.
Moreover, models and
observations of neutron-star low-mass X-ray binaries suggest that
accretion may cause the field to decay (see, e.g., Bhattacharya et
al.\ 1992), so the field strength during accretion was at least as
large as it is now.  Therefore, we can solve for the minimum
luminosity during accretion by setting $P_{\rm eq}=P$ and
substituting in $\mu=8.8\times 10^{26}$~G~cm$^3$ and $R_6=1$.
The result is $L_{37}>0.9(M/M_\odot)^{-2/3}(\omega_c/0.8)^{7/2}$.
Even for a relatively massive neutron star with $M=2\,M_\odot$, 
this is $5\times 10^{36}$~erg~s$^{-1}\approx 10^3L_\odot$.

At the distance of the inner planet, $r=3\times 10^{12}$~cm, the
equivalent blackbody temperature for a perfectly absorbing surface
is $T=[L/(\sigma 4\pi r^2)]^{1/4}=5300$~K even for $M=2\,M_\odot$.  
At the $\sim$keV energies
typical of accretion emission from a neutron star surface, the absorption
cross section is much greater than the scattering cross section, especially
if heavy elements are present, so the assumption of nearly perfect absorption 
is likely to be good.  Furthermore, the dependence on albedo is very weak
(only the 1/4 power), so the temperature estimate above is robust.
This temperature is sufficient to
boil any element likely to be in abundance around the star.  Therefore,
either direct evaporation or ablation 
due to the impinging X-rays can proceed efficiently.

Consider evaporation first, assuming that the innermost planet is
formed of an element of atomic weight $A$.  We also conservatively assume
a density $\rho\approx 10$~g~cm$^{-3}$; a more realistic, less dense
planet would be more easily evaporated.  
The radius of the planet is then $1.3\times 10^8$~cm, so the
scale height of the atmosphere created is $h=kT/mg=1.2\times 10^9A^{-1}$~cm.
If the planet is composed of elements significantly lighter than iron
($A=56$), then the scale height is comparable to $R$, so the illumination
will cause the planet to swell up and disperse on about a sound crossing time.
Even for iron, $h\approx 0.15\,R$.  The weakening of gravity with
increasing radius means that a substantial fraction of the gaseous iron
will be at large enough distances to escape; for example, a
calculation including the $r^{-2}$ dependence of gravitational acceleration
shows that $\approx 1$\% of the planet will be at radii $>4\,R$, and
$>$0.1\% will be unbound.  Therefore, even if the innermost planet is
pure iron, it will be evaporated in a few hundred sound crossing times,
a matter of only days.  Note that even if there is an optically thick disk
present, the blackbody temperature at the orbital radius of the planet is
unchanged, so evaporation will proceed efficiently.

A separate argument, which is also important for the two outer planets,
is that the X-ray illumination can ablate the planets directly.
Numerical analyses (e.g., Phinney et al.\ 1988; van den Heuvel et
al.\ 1988) suggest that a fraction $\sim$10\% of the luminosity
intercepted by a gaseous companion to a neutron star ablates that companion,
and that the material leaving the companion does so at approximately
the escape velocity.  A rough estimate of the time scale for
ablation therefore is obtained by dividing the gravitational binding
energy of the companion by 0.1 times the X-ray luminosity intercepted
by the companion.  If an optically thick exists at this time it will be
able to shield the planets from ablation.  However, as we show in the
next section, the required mass of such a disk is comparable to the mass
of the planets, and hence a disk massive enough to shield the planets
will drag them rapidly inwards as the disk accretes.
  
  At a distance of $3\times 10^{12}$~cm,
about $10^{-9}$ of the total luminosity falls on the surface of a
planet of radius $\sim 10^8$~cm, so
$10^{-10}$ of the total energy released by accretion is used for
ablation of the innermost planet.  
This total energy may be estimated from the amount of angular
momentum necessary to spin up the star to its current rotation frequency.
That frequency is $\omega\approx 10^3$~rad~s$^{-1}$, so for a neutron
star with
a typical moment of inertia $I\approx 10^{45}$~g~cm$^2$ the angular momentum
is $10^{48}$~g~cm$^2$~s$^{-1}$.  Assuming that the star was rotating
much more slowly than this prior to accretion, this is the amount of
angular momentum that must be accreted.  At the $\sim 5\times 10^6$~cm
radius at which the torque is exerted, the specific angular momentum
is close to its Newtonian form $\ell\approx\sqrt{GMr}$, or
$\ell\approx 3\times 10^{16}$~cm$^2$~s$^{-1}$ for $M=1.4\,M_\odot$.
The neutron star must therefore accrete $3\times 10^{31}$~g$\approx
0.01\,M_\odot$ to spin up to its current frequency.
Assuming that the accretion efficiency is $L/{\dot M}\sim 0.2\,c^2$,
typical for accretion onto the surface of a neutron star,
this will release a total of $6\times 10^{51}$~ergs, taking 
$4\times 10^7$~yr at $5\times 10^{36}$~erg~s$^{-1}$.  Multiplying
by $\sim 10^{-10}$ yields $6\times 10^{41}$~erg.
If we again conservatively assume that the companion is pure
iron, the binding energy at a radius $1.3\times 10^8$~cm 
is about $GM^2/R=4\times 10^{36}$~erg.  This is only $10^{-5}$ of the
ablation energy.  Therefore, even if evaporation
were somehow suppressed, the inner planet would
be ablated within a few hundred years, much shorter than the $\sim
10^{7-8}$~yr spent in the luminous LMXB phase.

The outer planets would also be in danger.  Their semimajor axes are
approximately twice that of the inner planet, so the temperature is
down by a factor $2^{1/2}$, to $\sim$3800~K.  This will boil any abundant
element except pure carbon, so again ablation would be highly efficient unless
the outer planets were truly diamonds in the roughest of environments.  If 
the densities of the outer planets are the same as that of the inner planet,
their radii scale like $M^{1/3}$, so their binding energies scale like
$M^2/R=M^{5/3}$.  Their surface area goes like $M^{2/3}$, so the ratio
of binding energy to intercepted luminosity, assuming the same radiation
flux, goes like $M$.  At twice the distance and $\sim$200 times the
mass of the inner planet, the ratio of binding energy to intercepted
luminosity is a factor $\sim$1000 larger for the outer planets, which
means that they will be evaporated in a time $\sim 10^5$~yr, still a
factor $\sim 100$ shorter than the accretion time.
Therefore, all three planets would be destroyed if they
were in their current positions and the pulsar was spun up by accretion.
This shows that if the pulsar was spun up by accretion, that accretion
had to take place before planets existed in the system.

These constraints are summarized in Figure~1.  Here we plot contours
of constant destruction time (from the combined effects of evaporation
and ablation) against the mass of the object
and the photon flux received, from $10^0$ years (leftmost contour) to
$10^7$ years (bottom right contour).  To give a conservative upper limit
to the destruction time we assume planets made of pure iron with densities
of 10~g~cm$^{-3}$.  The locations in this plot of
the three planets in the PSR~1257+12 system are indicated with dots.
For high fluxes, direct evaporation destroys the planet quickly, but
for lower fluxes ablation becomes more important, hence the change in
the slopes of the lines around a destruction time of $\sim 10^5$~years.
At a flux less than $F\approx 5\times 10^9$~erg~cm$^{-2}$~s$^{-1}$,
so that iron remains liquid, destruction by either evaporation or ablation
is highly inefficient.

\begin{figure}
\vskip -0.5truein
\vbox{\hskip 0.5truein
\psfig{file=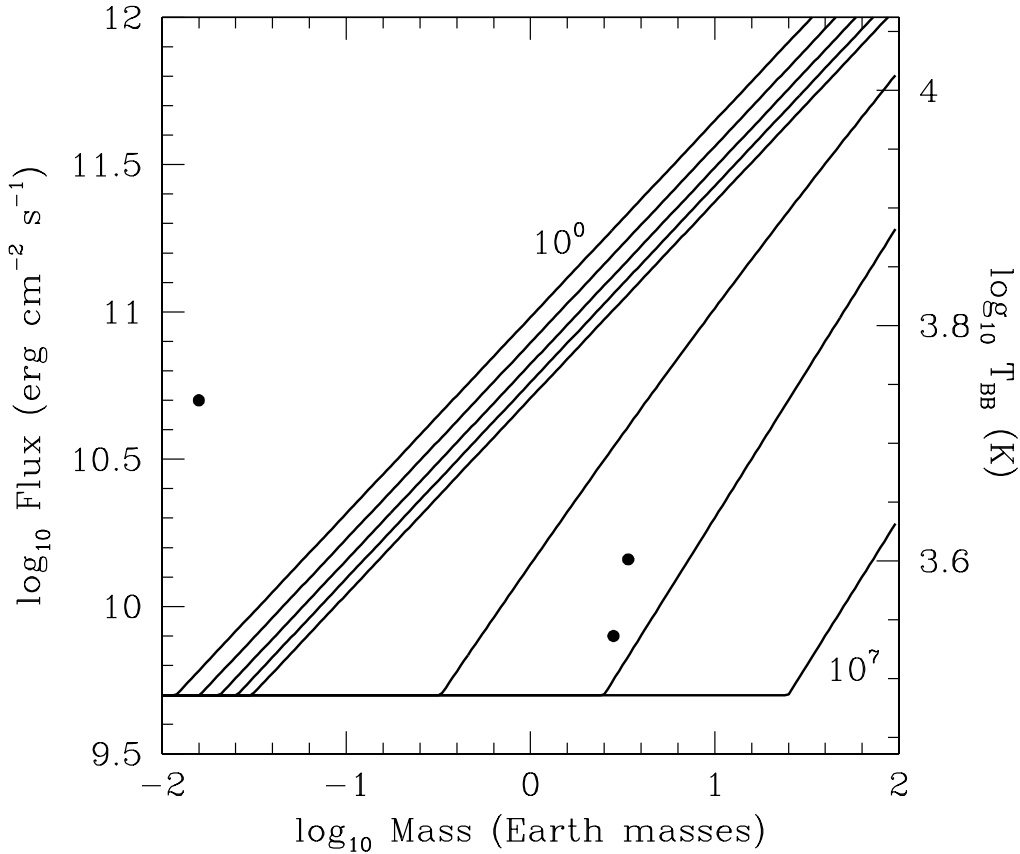,width=5.5truein,height=5.5truein}}
\caption
{Destruction time as a function of planetary mass and photon flux.
  The equivalent blackbody temperature is plotted on the right hand
  axis.  As discussed in \S~2.2, if the illuminating flux due to
  accretion luminosity is high enough to make a planet gaseous, either
  evaporation or ablation can destroy the planet efficiently.  This
  figure shows the destruction time from the sum of these two
  processes, versus log flux and log mass of the planet.  For
  comparison, the solar flux at Earth is $\approx 1.3\times
  10^6$~erg~cm$^{-2}$~s$^{-1}$.  The time contours are spaced by one
  order of magnitude each in years, from $10^0$ (far left) to $10^7$
  (far right).  We assume conservatively that the planets are pure
  iron, with a density of 10~g~cm$^{-3}$.  More realistic compositions
  would be less dense and hence easier to destroy.  Below a flux of
  $5\times 10^9$~erg~cm$^{-2}$~s$^{-1}$ ($T_{BB} = 3000$K) iron is
  molten, not gaseous, so evaporation and ablation are negligible.  At
  very high fluxes the scale height of the atmosphere is significant
  compared to the planetary radius, so evaporation proceeds rapidly.
  This is the case for the contours from $10^0$ years to $10^4$ years.
  At lower fluxes, evaporation is inefficient and ablation takes over,
  hence the different slopes of the $10^6$ year and $10^7$ year
  contours.  The slight change in slope in the $10^5$ year contour
  results from the transition between the evaporation and ablation
  regimes.  The current locations of the three planets in the
  PSR~1257+12 system are shown with dots, assuming inclinations of
  $\sin i=1$.  Clearly, the innermost planet is well into the
  evaporation regime whereas the two outer planets are in the ablation
  regime.}
\end{figure}

\subsection{Effect of pulsar particle luminosity}

   The previous section concentrated on the effect of photon luminosity
on already-existing planets.  Now consider the effect of particle luminosity
on a protoplanetary disk.  This is of relevance to models of the PSR~1257+12
system such as that of Banit et al.~(1993), in which planets form from
matter ablated from a companion star.  It is also important in projecting
whether any millisecond pulsars are expected to have planetary systems
consisting solely of small, asteroid-sized, objects.  We show that the
particle luminosity is great enough that if there is a time in the evolution
of the system when a massive disk does not exist, then a massive disk cannot 
form because it would be ablated efficiently by the particle luminosity.  

The spindown energy of isolated pulsars is
thought to be released primarily in the form of highly relativistic
particles.  These particles interact with matter in a fundamentally
different way than do X-rays.  X-rays interact with electrons, and are
not energetic enough to eject nuclei in one collision.  In contrast,
the relativistic particles are typically nuclei, which interact with
other nuclei.  Their energy is enormous compared to any relevant binding energy
(gravitational or chemical), and hence can eject a particle in a single
collision.  Therefore, whereas X-ray ablation has a threshold in intensity,
ablation by relativistic particles always occurs, albeit at negligible
rates in some cases.

If the disk is optically thin to these particles, then only
a small fraction of the spindown energy is imparted to the disk.  Any
impacts will deposit far more than the specific binding energy, and hence
any nucleus in the disk hit by a high-energy particle will be sent to
infinity, but the energy per particle will be so high that the resulting
mass flux from the disk will be small.  If we now
imagine that the optical depth of the disk is increased, then a larger and
larger fraction of the spindown energy is transferred to the disk.  If the
disk is still optically thin to the high-energy particles then each high-energy
particle only interacts with one disk nucleus, and the mass flux will scale
roughly with the optical depth of the disk.  However, when the optical depth
to the high-energy particles exceeds unity, then each high-energy particle
from the neutron star will interact with more than one target nucleus, and
the collisions will spawn further high-energy particles that interact in
their turn.  The energy imparted per particle is therefore diminished,
meaning that the mass flux from the disk (which depends on the total energy
deposited in the disk divided by the average energy per ejected particle)
is increased more rapidly than
the optical depth.  If, however, the optical depth is very large, then the
average energy per particle drops below the gravitational binding energy, 
and the mass flux drops quickly.  The maximum mass flux therefore occurs
when the average energy per particle is comparable to the binding energy.

   Similar considerations apply to the formation of a planet from grains
in such an
environment, assuming that the grains are unshielded from the pulsar particles.
Small grains are affected little, because they are optically thin.  By
contrast, larger rocks (with optical depth slightly larger than unity)
are exposed to the full flux of the particles, and again the mass flux
produced by the particles peaks when the average energy per particle
is comparable to the binding energy, although here the binding energy is 
chemical instead of gravitational.

   If there is a supply of matter to a nascent disk (as in the picture
of Banit et al.\ 1993, in which the supply comes from the ablation of
the stellar companion), then the mass of the disk depends on whether the
supply rate ${\dot M}_{\rm supply}$ is greater than or less than the
maximum rate ${\dot M}_{\rm max}$ at which ablation can remove mass.
If ${\dot M}_{\rm supply}>{\dot M}_{\rm max}$ then the disk increases
steadily in mass, indefinitely in principle.  If 
${\dot M}_{\rm supply}<{\dot M}_{\rm max}$ then the disk will reach
an equilibrium mass when ablation balances supply.

   Let us now quantify this picture.  Ultrarelativistic particles 
interacting with protons have a stopping column depth of about
$\sigma_p$=100~g~cm$^{-2}$ (e.g., Slane \& Fry 1989).  
At the distance $R=3\times 10^{12}$~cm of
the inner planet in the PSR~1257+12 system, the gravitational binding energy
is about $U=10^{14}$~erg~g$^{-1}$, and the area of a sphere at radius $R$
is $A\approx 10^{26}$~cm$^2$.  The current spindown luminosity of the
pulsar is ${\dot E}\approx 10^{34}$~erg~s$^{-1}$.  If the disk subtends
a solid angle that is $\epsilon=0.1$ of the whole sphere, this means that
the maximum mass flux in the wind is ${\dot M}_{\rm max}=
\epsilon{\dot E}/U=10^{19}$~g~s$^{-1}$.  To estimate ${\dot M}_{\rm supply}$
we assume a supply of matter due to evaporation of a stellar
companion (as in the ``black widow" pulsar PSR~1957+20).  We also make the
generous assumption
that the evaporation rate is equal to that in the black widow pulsar even
though the spindown luminosity of PSR~1957+12 is ten times less than that
of PSR~1957+20.  Then ${\dot M}_{\rm supply}=10^{17}$~g~s$^{-1}$.
This is less than ${\dot M}_{\rm max}$ by two orders of magnitude, and
hence an equilibrium disk mass will be reached.  Assuming 
that equilibrium occurs at an optical depth of a few, the total 
mass in equilibrium is about $10^{27}$~g.
In addition, since the mass is coming from the companion,
about 97\% of this mass is expected to
be in hydrogen or helium, leaving no more than $10^{26}$~g and probably
a factor of a few less in metals, compared with the $\sim 4\times 10^{28}$~g in
the two largest planets in the system.  Moreover,
relativistic particles would shatter complex 
nuclei and reduce the metal fraction even more.  The remaining mass
in metals would be far less than needed to form the current planets.

   We now consider whether planets can form in this environment from
small grains, if the grains are not shielded from the pulsar radiation.
The relevant binding energy $U$ is molecular,
which we assume is $\sim$1~eV per molecule or about
$10^{11}$~erg~g$^{-1}$.  The particle flux at the distance of the
innermost planet is $F\approx 10^{34}$~erg~s$^{-1}/10^{26}$~cm$^2
=10^8$~erg~cm$^{-2}$~s$^{-1}$.  The mass loss rate from
the rock due to radiation therefore has a maximum ${\dot m}_{\rm
max}=FA/U$, where $A=\pi a^2$ is the cross-sectional area of the rock.
For a stopping column depth of 100~g~cm$^{-2}$, the radius of a grain
with optical depth $\tau\sim 10$ (so that the average energy per particle
is close to the molecular binding energy) is $a\sim 100$~cm.
The timescale for evaporation of a particle of mass $m={4\over 3}\pi
a^3\rho$, where $\rho$ is the density, is then $t_{\rm evap}=m/{\dot
m}={4\over 3} \rho aUF^{-1}$.  This is $<10^6$~s for $a<100$~cm and
any reasonable
density.  Therefore, in this environment rocks cannot grow from small
grains to radii greater
than a few tens of centimeters.  If larger
planetesimals are present in the system prior to exposure to the particle
radiation, then the radiation will tend to ablate the planetesimals.  We
may estimate the rate of ablation by computing the area strongly affected
by the high-energy particles.  This area is the region with a path length
through the planetesimal such that the energy per particle at exit
exceeds the chemical binding energy.  Calling this distance $d$ and the
planetesimal radius $R$, the cross-sectional area with a path length less
than $d$ turns out to be simply ${1\over 4}\pi d^2$, independent of $R$
if $R>d/2$.  The mass
loss rate ${\dot m}$ is therefore also independent of $R$, and hence the
evaporation time scales as $m\sim R^3$.  Hence, within $10^{7-8}$~yr all
planetesimals of radius less than $\sim 1$~km will be evaporated by the
$10^{34}$~erg~s $^{-1}$ flux from PSR~1257+12.  The lower limit to the
survival radius may be even larger, because the main face of a
planetesimal with radius 1~km has an area a factor of a million larger
than $\pi d^2$, with $d\approx 100$~cm.  A small fraction of this incident
radiation may ablate additional matter.  The lower limit to planetesimal
radius is thus probably $\sim 1-10$~km.

   Particle radiation therefore prevents the formation
of planets when the disk is not very optically thick to the high-energy
particles.  Thus, for planets to form around a pulsar, the disk
must have sufficient mass to shield itself from the radiation.  That is,
the energy of the relativistic particles must be degraded sufficiently
that their average energy is less than the binding energy.  In a single
collision, the most efficient reduction of energy occurs when both the
primary nucleus and the target nucleus receive half of the original energy.
One can therefore conservatively assume that in each interaction, the
maximum energy per particle is reduced by a factor of two, so that after
$n$ scatterings the typical energy after propagation
through the medium is reduced by a factor $\sim 2^n$.  As long as the
particles are relativistic, forward beaming means that the number of
scatterings after traversing an optical depth $\tau$ is $n\sim\tau$.
When the particles are nonrelativistic then beaming is minor, and hence
the number of scatterings increases more rapidly, as $n\sim \tau^2$.
Assuming an initial Lorentz factor of $\sim 10^{3-5}$, the minimum optical 
depth required to reduce the particles to nonrelativistic energies is 
therefore conservatively $\tau\sim 10-20$.  Assuming a disk covering fraction 
of $\epsilon\sim$10\%, the required disk
mass is then at least $M_{\rm tot}=\epsilon \pi r^2\tau\sigma_p$, or about
$1-2\times 10^{28}$~g at a radius $r=3\times 10^{12}$~cm.  If the initial 
disk mass
is less than this, we expect no planets to form.  Given that this is already
greater than one Earth mass, this may mean that isolated millisecond pulsars 
either have planetary-mass objects around them or nothing, and hence that
we should not expect systems with just asteroids.  If so, it suggests that 
microsecond timing noise in millisecond pulsars is not dominated by
asteroids, as was discussed as a concern by Wolszczan (1999).

\subsection{Constraints from the lack of planets around other isolated MSP}

In addition to the constraints just listed, formation mechanisms for
planets around millisecond pulsars have another constraint, common to
scenarios proposed for any rare object: the mechanism cannot be {\it too}
good, or else more examples would be seen.  In this case, why are there
no other planets around isolated MSP, given the extreme sensitivity of
MSP timing to such perturbations?  Of course, it could be that the 
PSR~1257+12 system had a unique history, but here we make the Ockham's
razor assumption that all isolated millisecond pulsars are formed in
the same way.

Of the nine isolated MSP in the Galactic disk, only
PSR~1257+12 has confirmed planets around it, even though asteroid-sized
objects with orbital periods of a few years or less could be detected with
current techniques (Wolszczan 1999).  The gap of at least three orders of
magnitude between these mass upper limits and the mass of the planets
around PSR~1257+12 suggests that the pulsar planetary system is not
simply in the high-mass tail of a distribution, but is instead the result
of a rare event.  This argues against disrupted companion scenarios
(Podsiadlowski et al. 1991; Fabian \& Podsiadlowski 1991), in which there
is always $0.01-0.1\,M_\odot$ in the disk after disruption.  Such a high
disk mass is expected to be extremely favorable for the formation of planets,
because the column depth is high enough to shield the matter from the
pulsar flux (see above).  It would therefore be surprising that only one
MSP has planetary-mass objects around it.  Instead, an idea such as one
in which supernova recoil kicks the neutron star through the companion
(Phinney \& Hansen 1993), has
many desirable properties.  In this picture, only if the neutron star
intersects the companion will it accrete mass and potentially form planets.
Otherwise, virtually no mass is accreted and the star simply spins down
in isolation.  We explore this idea further in the next section.

\section{Allowed Formation Histories}

The physical constraints in \S~2 may be summarized
as follows.  (1)~If PSR~1257+12 was spun up by accretion from a
companion, then ablation makes formation of the planets 
before or during this accretion highly implausible.
(2)~Particle radiation from the pulsar
will destroy a disk if the disk has too low a mass.  It will also prevent
large grains from forming in an unshielded environment.  Therefore, either
supply of mass to the disk must exceed the mass loss rate or the disk mass
must initially be well in excess of $\sim 10^{28}$~g.  (3)~The formation
mechanism cannot be inevitable for isolated millisecond pulsars, or other
examples would be seen.  

The most plausible mechanisms are therefore those in which an isolated 
neutron star sometimes (in $\sim$10\% of cases) obtains a disk of mass 
$>10^{28}$~g from which planets form, but in most cases does not acquire 
significant mass.  This favors ideas such as the supernova recoil scenario
(Phinney \& Hansen 1993).  Given that
1 out of 9 isolated MSP have planets, then prior
to the supernova the stellar companion must subtend a few percent to a
few tens of percent of the
sky, assuming that the direction of the kick delivered to the neutron
star is not correlated with the direction to the companion.  This would
require a separation of a few times the radius of the presupernova star,
if the stellar companion is also a massive star, implying a separation 
of $\sim 10^{12}-10^{13}$~cm.  The observed distribution of initial orbital
separations $a_{\rm init}$ of massive stars is $\sim 1/a_{\rm init}$
(e.g., Kraicheva et al.\ 1979), so tens of percent of massive binaries
are expected to be in this range of separations.

If the neutron star receives a kick in the direction of the companion,
then in order to eventually form planets it needs to capture at least
$10^{28}$~g from the companion.  We can make a very rough estimate of
the mass captured by making a Bondi-Hoyle type assumption that the
matter captured from the companion star has an impact parameter $b$
relative to the neutron star that is less than $b_{\rm max}$, where
$b_{\rm max}$ is defined such that the effective orbital velocity
$v_{\rm orb}=(GM_{\rm NS}/b_{\rm max})^{1/2}$ is equal to the kick
velocity $v_{\rm kick}$ of the neutron star. The proper motion of
PSR~1257+12 is measured at 300~km~s$^{-1}$ (Wolszczan 1999);
Identifying this as the kick velocity yields $b_{\rm max}\approx
10^{11}$~cm. If this matter is captured with an efficiency
$\epsilon\approx 10^{-3}-10^{-1}$ then the amount of matter captured
from the companion is typically $M_{\rm cap}\sim\epsilon b_{\rm max}^2
R_c{\bar\rho}_c$, where $R_c$ and ${\bar\rho}_c$ are the radius and
average density of the companion.  For a companion of mass
$10\,M_\odot$, radius $R_c=10^{12}$~cm, and average density
$\rho_c=10^{-2}$~g~cm$^{-3}$, the captured mass is $\approx
10^{29-31}$~g, which is in the needed range. Although the initial
scale of the disk, $b_{\rm max}\approx 10^{11}$~cm, is two orders of
magnitude smaller than the current planetary system, conservation of
angular momentum implies that the disk will spread significantly as it
evolves. Planets will form only after the outermost portions of the
disk have spread and cooled enough to allow efficient condensation of
solids (Lissauer 1988). Thus planetary systems formed by this
mechanism should consist of terrestrial mass planets confined within a
few AU of the central pulsar.

Such a disk mass would effectively shield the protoplanets from the
radiation of the neutron star.  In addition, when particle radiation
dominates the emission from the pulsar (as opposed to accretion
radiation) the luminosity of $\sim 10^{34}$~erg~s$^{-1}$ would produce
a blackbody temperature of only a few hundred Kelvin at the distances
of the planets.  At such temperatures, ionization is low and molecule
and grain formation is thought to proceed efficiently (Lissauer 1993).
This is therefore an environment supportive of planet formation, particularly
given that the characteristic age $\tau_c\sim 10^9$~yr of the pulsar
is much longer than the $\sim 10^7$~yr required for planet formation.  The
allowed mass range of the disk would also accommodate fourth planet
with a mass $\simless 100\,M_\oplus$ if the existence of this planet
is confirmed by a long baseline of observations.

This scenario implies that many isolated neutron stars may have planets
in orbit around them.  Why, then, are there no other known planets
around isolated pulsars, millisecond or otherwise?  There is a reported
detection of a planet in the PSR~B1620-26 system, but this is a triple
system in a globular cluster and probably formed by a different
mechanism, such as an exchange interaction (Sigurdsson 1993;
Ford et al.\ 2000).  We
attribute the paucity of detected pulsar planets to the stringent
conditions for such planets to be observable.  Young radio pulsars such
as the Crab or Vela are extremely powerful emitters of particle
radiation. For example, the spindown luminosity of the Crab pulsar,
which is thought to emerge primarily as high-energy particles, is $\sim
5\times 10^{38}$~erg~s$^{-1}$.  This will prevent planetary formation
within several AU, and will continue to do so until the spindown
luminosity drops by more than an order of magnitude.  In addition, the
formation of planetary-mass objects is thought to require $\sim
10^7$~yr, by which point any pulsar with a magnetic field $\sim
10^{12}$~G or higher will have spun down past the death line, and will
therefore not be detectable as a pulsar.  As a consequence, planets
would not be detectable around the neutron star.  Finally, even if
planets are present around a young, high-field pulsar, the timing
noise makes difficult the  detection of sub-Earth mass planets
(Wolszczan 1999).

Hence, if an isolated neutron star has planets around it, the only
chance for their presence to be detected is for the neutron star to
have a weak magnetic field, so that it remains a pulsar long enough for
planets to form.  For the pulses to be detected, the spin frequency 
then has to be high, otherwise the total spindown energy is too low.
This means that the star is born with a high frequency, and that accretion
from any remnant disk does not spin the star down significantly.
With these constraints on observability, it is not surprising that
there is only one known planetary system around an isolated pulsar.

\section{Implications for Millisecond Pulsars}

Our picture requires that the neutron star in PSR~1257+12 was born
with approximately its current magnetic field strength of $B\approx
10^9$~G.  If instead it were born with a $\sim 10^{12}$~G magnetic field then
it would have spun down rapidly to a period of order seconds, either because
of magnetic dipole spindown or because of accretion torques. To be
spun up by accretion to millisecond periods, the accretion would have
had to proceed at near-Eddington rates after the field had decayed to
$B\sim 10^9$~G.   The decay time is at least $10^7-10^8$
years for solitary pulsars  (Bhattacharya et al. 1992). Moreover, the
existence of neutron stars with $B\sim 10^{12-13}$~G in high-mass
X-ray binaries, which often accrete at near-Eddington rates and have
accretion lifetimes of millions of years, suggests that even active
accretion does not cause the field to decay in less than $\sim
10^6$~yr.  Therefore, the accretion rate would have had to be close to
Eddington after several million years.  Studies of accretion from a
remnant disk (Cannizzo, Lee, \& Goodman 1990) show that the system
becomes  self-similar quickly and that the accretion rate drops like a
power law in time, ${\dot M}\propto t^{-7/6}$, so such a high
accretion rate millions of years after the onset of accretion would
imply an unrealistically high initial accretion rate.

We also suggest that the initial spin frequency of the neutron star
was close to its current value.  Otherwise, the star would
have to be spun up by accretion.  Even if the initial field was weak,
for accretion from the disk to spin the star up to its current rate
would again require near-Eddington accretion for several million
years, and the low accretion rate tail of the accretion from the disk
is likely to
slow the star down below its current spin rate.  Therefore, we propose
that the pulsar in this system, and by extension perhaps all isolated 
millisecond pulsars, was born with approximately its current spin rate and
magnetic field strength.  We emphasize that these considerations need
not apply to isolated millisecond pulsars in globular clusters, in which
there are other possible 
formation channels for isolated MSP (e.g., exchange interactions;
Sigurdsson 1993).

The link between millisecond pulsars and neutron star low-mass X-ray
binaries is well established.  Some 80\% of millisecond pulsars in the
Galactic disk are
in binaries, compared to only 1\% of slower pulsars (Lorimer 2000).  In 
addition, the recent discovery of 401~Hz pulsations from the LMXB 
SAX~J1808--3658 (Wijnands \& van der Klis 1998) is the strongest evidence
yet that neutron stars in LMXBs can have spin frequencies comparable to
those of MSP.  It was thought a decade ago (Kulkarni \& Narayan 1988)
that there was a significant discrepancy between the birthrate of
short orbital period ($<$25 day) MSP and their presumed progenitor
LMXBs, in that the pulsar birth rate was two orders of magnitude too
large.  However, better statistics have been collected and the factor
is now less than four (Lorimer 2000).

Isolated millisecond pulsars may pose a different problem, however.
The small number of isolated MSP makes sample variance a significant
concern, but analysis of the current data
suggests that isolated MSP have a lower luminosity than MSP in binaries,
and may have other distinct properties as well (Bailes et al.~1997).
In addition, the best estimates of the birthrate of isolated MSP in
the Galaxy, $2\times 10^{-5}$~yr$^{-1}$ (Lorimer 2000), are at least 
ten times the
birthrate estimates for binary MSP (note, however, that the isolated MSP
estimate is strongly influenced by the large weights attached to
a few low-luminosity sources).  If this rate and these discrepancies
are confirmed by future surveys with larger samples, it suggests that
isolated millisecond pulsars are formed by a different channel, which
does not involve recycling (Bailes et al.\ 1997).

Our analysis of PSR~1257+12 suggests that the different channel may
simply be that some neutron stars are born with fast spins and weak
magnetic fields.  Birth with rapid spin is compatible with the
probable origin of the Crab pulsar and other similar pulsars, which
could have been born with millisecond periods.  Birth with a weak field
is not directly comparable to the known population of young pulsars, although
with a birth rate of $2\times 10^{-5}$~yr$^{-1}$ in the Galaxy and assuming
a supernova remnant lifetime of $\sim 10^5$~yr it is not surprising
that no known supernova remnant harbors a millisecond pulsar.  It is
therefore possible that pulsars born with weak fields are simply at the
tail end of a distribution peaked at $B_{\rm init}\sim 10^{12}$~G, and
it is equally possible that weak-field isolated pulsars are the product
of a completely different set of processes in the supernova, or of
rare types of supernovae.

\section{Conclusions}

We have argued that the physical constraints on the PSR~1257+12
system, in particular the existence of the small innermost planet,
in addition to the lack of planets around other isolated millisecond
pulsars, points strongly towards an evolutionary history in which
the neutron star had a high initial spin rate and weak initial magnetic
field and formed planets from a disk of captured matter.  We also
predict that in the absence of other planets, objects of asteroid
size or smaller will not form around millisecond pulsars, due to ablation
by the flux of high-energy particles.

\acknowledgements
We dedicate this paper to the memory of John Wang, a valued friend and
colleague and an outstanding astrophysicist.  This work was supported
in part by NSF Career Grant AST9733789 (DPH) and NASA grant NAG 5-9756
(MCM).

\end{document}